\documentclass{jpp}
\usepackage{graphicx}
\usepackage{epstopdf, epsfig}
\usepackage{hyperref}

 \def\bl{\par\vskip 12pt\noindent}
\shorttitle{Hall instability}
\shortauthor{Leonid Kitchatinov}

\title{Hall instability: origin, properties, and asymptotic theory
for its tearing mode}

\author{Leonid Kitchatinov \aff{1,2}
  \corresp{\email{kit@izf.irk.ru}}}

\affiliation{
\aff{1}
Institute of Solar-Terrestrial Physics SB RAS,
Lermontov Str. 126A, 664033, Irkutsk, Russia
\aff{2}Pulkovo Astronomical
Observatory, St. Petersburg, 196140, Russia
}

\begin{document}

\maketitle

\begin{abstract}
Hall instability in electron magnetohydrodynamics is interpreted as the
shear-Hall instability driven jointly by helicoidal oscillations and
shear in the electron current velocity. This explanation suggests an
antiparallel orientation of the background magnetic field and vorticity
of the current velocity as the necessary condition for Hall instability.
The condition is tested and generally confirmed by numerical
computations in plane slab geometry. Unstable eigenmodes are localized
in the spatial regions of the antiparallel field and vorticity.
Computations of the tearing-type mode of the instability are
complemented by (and generally agree with) asymptotic analytical
estimations for large Hall numbers. The stabilizing effect of
perfect conductor boundary conditions is found and explained. For large
Hall numbers, the growth rates approach the power law dependence $\sigma
\propto B^\alpha\eta^{1-\alpha}$ on the magnetic field ($B$) and
diffusivity ($\eta$). Almost all computations give the power index
$\alpha = 3/4$ with one exception of the tearing-type mode with vacuum
boundary conditions for which case $\alpha = 2/3$.
\end{abstract}

\section{Introduction}
The Hall effect is studied mainly in relation to its ability to catalyze
the dissipation of magnetic fields. This effect conserves magnetic
energy and magnetic helicity but it can amplify resistive dissipation by
decreasing the spatial scale of magnetic structures. \citet{GR92} noted
that hypothetical Hall turbulence can accelerate the decay of magnetic
fields in a neutron star crust towards their observationally detected
rates; the Hall cascade has been studied numerically by \citet{HR02} and
\citet{B20}. \citet{VCO00} considered a similar effect for the solar
corona.

Alternatively, dissipative magnetic structures can be produced by
instabilities. Growth rates of hydromagnetic tearing instability are
known to increase with allowance for the Hall effect \citep[see][for
recent publications]{ZMW17,Sea20,BBS21}. This is probably because the
Hall effect can drive its own instability of the tearing type.

Hall instability was discovered by \citet{RG02}. They found that the Hall effect can destabilize an equilibrium magnetic field if second-order spatial derivatives of the field do not vanish. The instability requires inhomogeneity in the background state for its development but
can onset under diverse conditions. If the background state includes
reversals of magnetic field, the instability can operate in the tearing
mode \citep{GH16}. Reversals are however not necessary and an
inhomogeneous background field without reversals can be unstable as well
\citep{RG02,PG10}. Inhomogeneity of density also can facilitate
the instability \citep{RKG04,WHL14}. It is not clear at the moment
whether the instabilities driven by the Hall effect under these diverse
conditions are of the same origin. There is no commonly accepted
pictorial explanation of Hall instability/instabilities.

A possible explanation was proposed in the preceding paper
\citep{Kit17}. The explanation relies on analogy with shear-Hall
instability \citep[cf.\,Sect. 8.4.3 in][]{RH04,Kunz08} and suggests that the Hall
instability results from shear in the electron current velocity. This
paper tests the explanation by computations of the Hall instability in
the electron MHD approximation in plane pinch geometry.

The next Section 2 briefly reminds the essence of the Hall effect and
the proposed explanation for Hall instability. Section 3 describes the
design and equations of the model. Section 4 presents the results of
numerical and analytical estimations for the Hall instability of the
background field profiles with and without reversal of the field
direction. The final Section 5 summarizes the results and concludes.

\section{Hall effect and Hall instability}
The Hall effect is related to the magnetically induced anisotropy of
electric conductivity. If the gyrofrequency $\omega_\mathrm{e}$ of
electrons - the main carriers of electric current - is not small
compared to the frequency $\nu_\mathrm{ei}$ of electron collisions with
particles of other species, the conductivity $\sigma_\|$ along the
magnetic field lines is larger than conductivities for other directions.
The Ohm's law then reads \citep[see, e.g.,][Chapt.11]{S13}
\begin{equation}
    \boldsymbol{j}' = \frac{\sigma_\|}{1 + R_\mathrm{H}^2}\left[
    \boldsymbol{E}' + R_\mathrm{H}\boldsymbol{\hat{b}\times E}' +
    R_\mathrm{H}^2\boldsymbol{\hat{b}}(\boldsymbol{\hat{b}\cdot E}')\right] ,
    \label{2.1}
\end{equation}
where $\boldsymbol{\hat{b}}$ is the unit vector along the magnetic field
$\boldsymbol{B}$, $\boldsymbol{j} = \mathrm{c}(\boldsymbol{\nabla \times
B})/(4\upi )$ is the current density, $\boldsymbol{E}$ is the electric
field, the primed notations mean that the equation is formulated for the
reference frame co-moving with the fluid mass velocity $\boldsymbol{u}$,
and
\begin{equation}
    R_\mathrm{H} = \omega_\mathrm{e}/\nu_\mathrm{ei}
    \label{2.2}
\end{equation}
is the Hall number. Equation (\ref{2.1}) can be reversed to express the
electric field in terms of the current density:
\begin{equation}
    \boldsymbol{E}' = \left( \boldsymbol{j}' -
    R_\mathrm{H}\boldsymbol{\hat{b}\times j}'\right)/\sigma_\| .
    \label{2.3}
\end{equation}
If equation (\ref{2.3}) is used instead of its isotropic counterpart
$\boldsymbol{E}' = \boldsymbol{j}'/\sigma_\|$ in the standard derivation
of the induction equation of MHD \citep[see, e.g.,][Chapter 4]{P79}, the
derivation yields
\begin{equation}
    \frac{\partial\boldsymbol{B}}{\partial t} =
    \boldsymbol{\nabla\times} \left[ ( \boldsymbol{u}_\mathrm{c} +
    \boldsymbol{u})\times\boldsymbol{B} - \eta\boldsymbol{\nabla\times
    B}\right] ,
    \label{2.4}
\end{equation}
where $\eta = \mathrm{c}^2/(4\upi\sigma_\|)$ is the magnetic
diffusivity,
\begin{equation}
    \boldsymbol{u}_\mathrm{c} = -\boldsymbol{j}/(e n_\mathrm{e})
    \label{2.5}
\end{equation}
is the current velocity, and the expressions for the gyrofrequency $\omega_\mathrm{e} = e B/(m_\mathrm{e}c)$ and conductivity $\sigma_\| = e^2 n_\mathrm{e}/(m_\mathrm{e}\nu_\mathrm{ei})$ in terms of
the microscopic plasma parameters are used. The Hall effect in MHD,
therefore, means an additional current velocity (\ref{2.5}) for the
magnetic field advection.

This effect for a uniform background field together with a (prescribed)
plane shear flow can drive an instability. The sufficient condition for
the shear-Hall instability is a counter-alinement of the magnetic field
and shear-flow vorticity \citep{RH04,RK05}. The
shear-Hall instability allows a clear pictorial explanation in terms of
the combined action of shear in the mass velocity and helicoidal
oscillations induced by the Hall effect \citep[see discussion of fig.8
in][]{Kunz08}.

As the current velocity $\boldsymbol{u}_\mathrm{c}$ enters equation
(\ref{2.4}) in the same way as the mass velocity $\boldsymbol{u}$, it
should lead to the same consequences. Instability can be expected from
shear in the current velocity if its vorticity is counter-aligned to the
magnetic field. This is the interpretation of the Hall instability
supposed in this paper. The interpretation can explain the diversity of
conditions for which the Hall instability is met. It is clear why the
magnetic field should depend at least quadratically on position for the
instability \citep{RG02}: otherwise the current velocity has no shear.
It is also clear why density inhomogeneity can be relevant.
Inhomogeneous electron density in Eq.\,(\ref{2.5}) can be the reason for
a shear. The magnetic field and current vorticity are counter-aligned
near the local maxima of the (unidirectional) magnetic field. This
explains why reversals of the field direction are not necessary for the
instability.

The difference with the shear-Hall instability however is that the
unstable background field is necessarily inhomogeneous and the current
shear does not change the magnetic energy. This results in the diffusive
character of the Hall instability. Its growth rates scale as
$B^\alpha\eta^{1-\alpha}$ with $0 < \alpha < 1$ for the large Hall
number.
\section{The model}
We adopt the electron MHD approximation that omits the mass velocity
$\boldsymbol{u}$ in the induction equation (\ref{2.4}). The
approximation is justified for the solid crust of neutron stars and
possibly for the case of large microscopic viscosity compared to the
magnetic diffusivity. The case of large magnetic Prandtl number is
characteristic of the hot plasmas of stellar coronae \citep[][Sect.
3]{BS05}. Normalized variables are used with distances normalized to the
characteristic scale $L$ of the background field, the field strength
normalized with the amplitude $B_0$ of this field, and time measured in
diffusive units $L^2/\eta$. With this approximation and normalization,
the induction equation retains only one governing parameter of the Hall
number (\ref{2.2}) that is now defined with (constant) amplitude $B_0$
of the background field:
\begin{equation}
    \frac{\partial\boldsymbol{B}}{\partial t} =
    \boldsymbol{\nabla\times} \left[ R_\mathrm{H}\boldsymbol{B\times}
    (\boldsymbol{\nabla\times B}) - \boldsymbol{\nabla\times B}\right] .
    \label{3.1}
\end{equation}

We use a Cartesian coordinate system and consider the stability of the
magnetic field that has only one non-zero $z$ component and varies with
coordinate $x$ only,
\begin{equation}
    \boldsymbol{B} = \left(0,0, B(x)\right).
    \label{3.2}
\end{equation}
The linear stability of this background field to small disturbances
$\boldsymbol{B}'$ is considered. In all cases where instability is
found, the most rapidly growing modes are uniform along the $y$-axis.
The (toroidal) $y$-component $b(x,z)$ and (poloidal) component
$\boldsymbol{\nabla\times}(\boldsymbol{\hat y}\, a(x,z))$ normal to this
axis are convenient to distinguish in such 2D disturbances to satisfy
automatically the condition $\boldsymbol{\nabla\cdot b} = 0$
($\boldsymbol{\hat y}$ is the unit vector along the $y$-axis)
\begin{equation}
    \boldsymbol{B}'(x,z) = \boldsymbol{\hat y}\, b(x,z) +
    \boldsymbol{\nabla\times}(\boldsymbol{\hat y}\, a(x,z)) .
    \label{3.3}
\end{equation}
The equation system for the small disturbances can be obtained by
linearization of Eq.\,(\ref{3.1}). Coefficients of the resulting
equations do not depend on time or $z$. An exponential dependence on
these variables $b,a \propto \mathrm{exp}(\sigma t + \mathrm{i}kz)$ can
therefore be assumed that leads to the eignvalue problem:
\begin{eqnarray}
    \sigma a &=& -\mathrm{i}kR_\mathrm{H}B(x)\,b - k^2 a + \frac{\partial^2
    a}{\partial x^2} ,
    \nonumber \\
    \sigma b &=& \mathrm{i}kR_\mathrm{H}\left[ B(x)\left(\frac{\partial^2
    a}{\partial x^2} - k^2 a\right) - \frac{\partial^2 B}{\partial x^2}
    a\right] - k^2 b + \frac{\partial^2 b}{\partial x^2} .
    \label{3.4}
\end{eqnarray}
The second derivative of the background field accounts for the shear in
the current velocity. The neglect of the diffusive decay of this field
is justified if the growth rates of instability are large, $\Real
(\sigma ) \gg 1$.

The equations (\ref{3.4}) are solved in a plane slab with two boundaries
imposed at $x = \pm 1$. Unless otherwise stated, the results to follow
correspond to the vacuum conditions at the slab boundaries,
\begin{equation}
    b = 0,\ \ \frac{\partial a}{\partial x} = \mp k a\ \ \mathrm{for}\
    \ x = \pm 1 .
    \label{3.5}
\end{equation}
A finite background field at the boundaries for this case can be
understood as imposed by external sources. The conditions for perfect
conductor boundaries,
\begin{equation}
    \frac{\partial b}{\partial x} = 0,\ \ a = 0\ \ \mathrm{for}\
    \ x = \pm 1 ,
    \label{3.6}
\end{equation}
is an alternative possibility.

Numerical computations were performed with a uniform finite-difference
grid of 750 grid-points. As the unstable eigenmodes can include a fine
spatial structure, a fourth-order accurate finite difference scheme was
used for spatial derivatives.
\section{Results}
The eigenmodes of Eq.\,(\ref{3.4}) include the well-known helicoidal
oscillations. No overstable solutions were found however. All the modes
with positive growth rates have zero imaginary part of their
eigenvalues.

According to the proposed explanation for the instability, it can be
expected for the antiparallel orientation of the background magnetic
field and vorticity of the current velocity (\ref{2.5}). For the 1D
profiles of Eq.\,(\ref{3.2}) this means the inequality
\begin{equation}
    B(x)\frac{\partial^2 B(x)}{\partial x^2} < 0 .
    \label{4.1}
\end{equation}
We proceed by testing this condition with computations for different
field profiles. The cases of symmetric and antisymmetric profiles about
the slab midplane can be distinguished.
\subsection{Antisymmetric field profiles}
\subsubsection{Numerical computations}
Antisymmetric profiles $B(x) = -B(-x)$ can support unstable modes of the
tearing-type. We first test the profile
\begin{equation}
    B(x) = \sin (-\upi x/2) ,
    \label{4.2}
\end{equation}
which satisfies the condition (\ref{4.1}).

\begin{figure}
    \centerline{\includegraphics[width = \columnwidth]{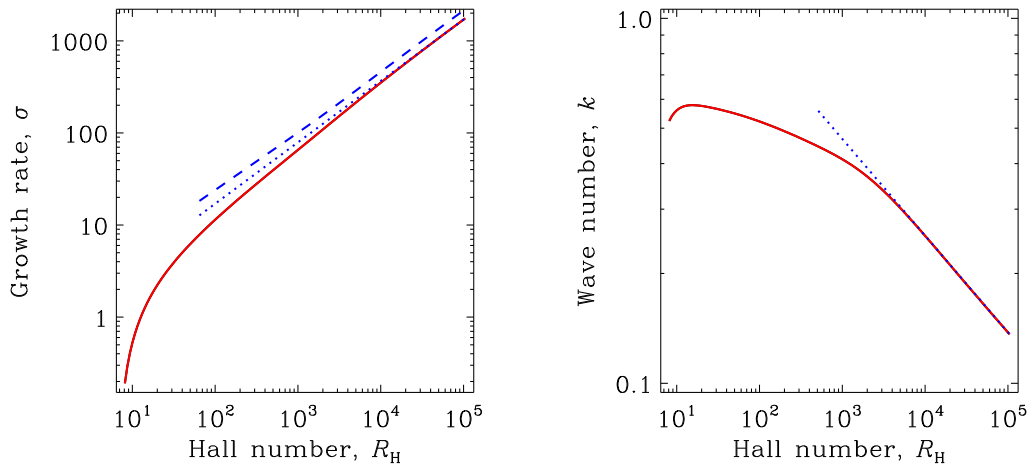}}
    \caption{{\sl Left panel}: Growth rate of the most rapidly growing mode for
           the field profile of Eq.\,(\ref{4.2}). The dotted line
           indicates the power law dependence $\sigma\propto R_\mathrm{H}^{2/3}$
           for large $R_\mathrm{H}$. The dashed line shows the result of analytical
           asymptotic estimation of the next Section. {\sl Right panel}:
           The wave number for which the
           maximum growth rate is attained. The power law approximation
           $k \propto R_\mathrm{H}^{-0.265}$ is shown by the dotted line.
           All for the vacuum boundary condition (\ref{3.5}).}
    \label{f1}
\end{figure}

Figure\,\ref{f1} shows the growth rates of the most unstable modes
and the wave numbers for which these maximum growth rates are obtained.
For a large Hall number, the growth rate approaches the power law
$\sigma\propto R_\mathrm{H}^{2/3}$ in agreement with \citet{GH16}.

\begin{figure}
    \centerline{\includegraphics[width = \columnwidth]{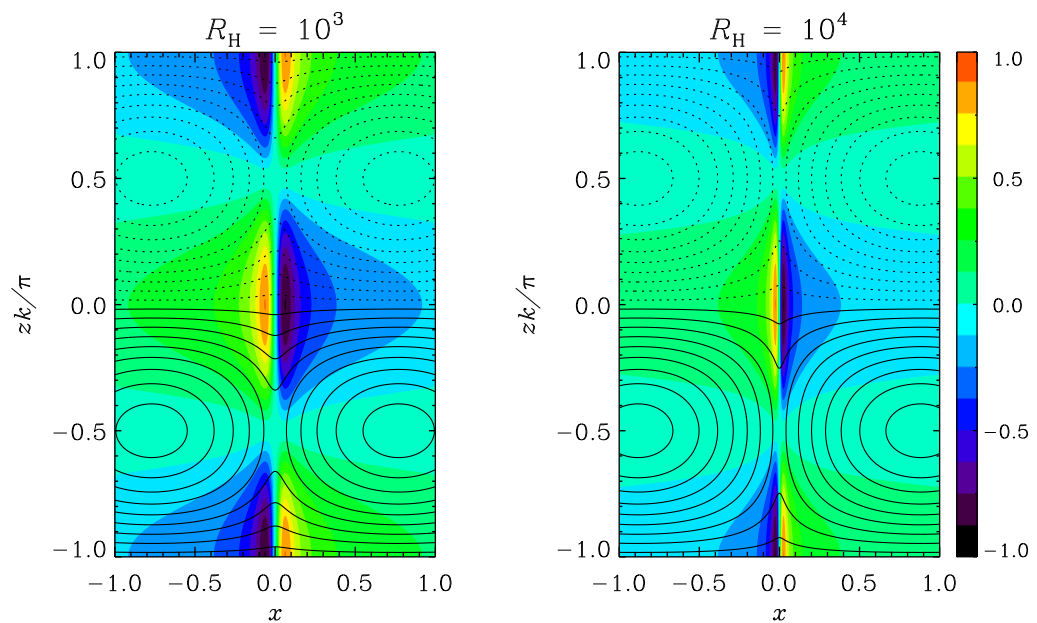}}
    \caption{Poloidal field lines and isolines of the toroidal field of most rapidly
        growing modes for $R_\mathrm{H} = 10^3$ ({\sl left panel}) and $R_\mathrm{H} = 10^4$
        ({\sl right panel}). Full and dashed lines correspond to the clockwise and
        anti-clockwise circulation of the poloidal field vector respectively.
        All for the field profile of Eq.\,(\ref{4.2}).}
    \label{f2}
\end{figure}

Figure\,\ref{f2} shows the eigenmode structure for two large values of
the Hall number. Reconnection of the field lines in a region near the $x
= 0$ position of the background field reversal is evident from this
figure. The thickness of the reconnection layer decreases with the Hall
number. The wave numbers of Fig.\,\ref{f1} are smaller than one.
Antisymmetric field profiles can be unstable to the long wavelength
tearing-type disturbances \citep{GH16}.

\begin{figure}
    \centerline{\includegraphics[width = 9 truecm]{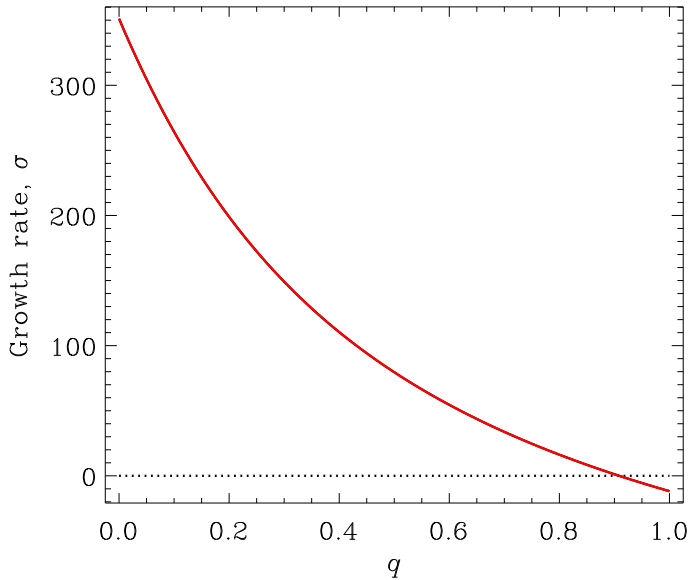}}
    \caption{Dependence of the growth rate on the $q$-parameter of the profile
         $B(x) = (1 - q)\sin(-\upi x/2) - q x$. $R_\mathrm{H} = 10^4,\ k = 0.25$.}
    \label{f3}
\end{figure}

Not all antisymmetric profiles are unstable however. Figure\,\ref{f3}
shows the growth rate as the function of the $q$-parameter of the
profile $B(x) = (1 - q)\sin(-\upi x/2) - q x$. The growth rate reverses
to negative values for $q \rightarrow 1$. The current velocity for the
linear profile of $q = 1$ is uniform and has no shear. The condition of
Eq.\,(\ref{4.1}) is then violated and the instability is switched off.

The local condition is also not sufficient for the global instability.
The instability is switched off with a change to the perfect conductor
boundary conditions (\ref{3.5}). This is probably because the helicoidal
oscillation takes part in the instability \citep{Kit17}. The helicoidal
rotation is blocked by superconducting boundaries.

\begin{figure}
    \centerline{\includegraphics[width = \columnwidth]{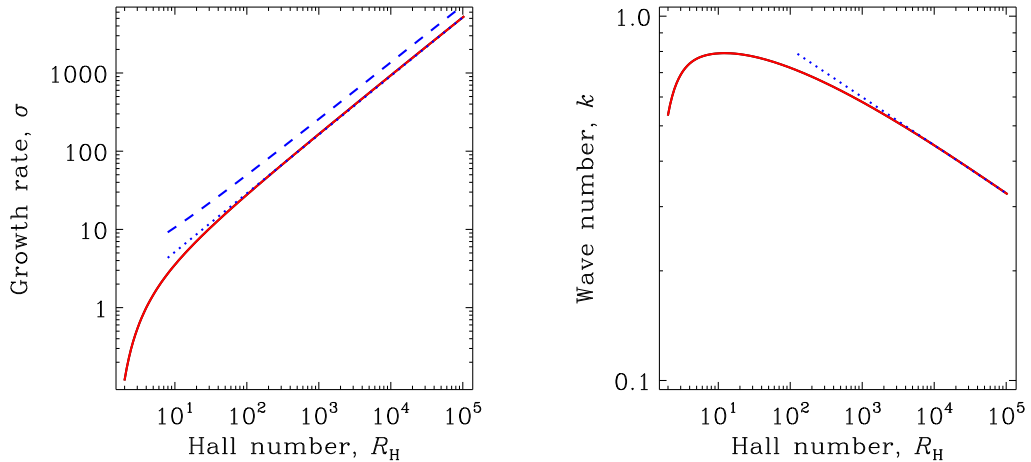}}
    \caption{The same as in Fig.\,\ref{f1} but for periodic boundary conditions of
           Eq.\,(\ref{4.3}). In this case, the asymptotic dependence of the growth
           rate on the Hall number changes to $\sigma\propto R_\mathrm{H}^{3/4}$.
           The dashed line shows the result of analytical asymptotic estimation.
           Scaling for the wave number also changes to $k \propto R_\mathrm{H}^{-0.132}$.}
    \label{f4}
\end{figure}

The growth rate scaling with the Hall number depends on the model design
also. \citet{Kit17} considered a periodic background field in unbounded
space. The profile of Eq.\,(\ref{4.2}) is mirror-symmetric about the planes of $x = \pm 1$ when considered in unbounded space. The eigenmodes for this periodic profile have to possess the same symmetry. This means that $x$-derivatives of the eigenfunctions
should be zero at $x = \pm 1$. The case of the periodic profile
(\ref{4.2}) in unbounded space can therefore be reproduced in the slab
model of this paper with the \lq periodic' boundary conditions
\begin{equation}
    \frac{\partial a}{\partial x} = \frac{\partial b}{\partial x} = 0\ \
    \mathrm{for}\ \ x = \pm 1 .
    \label{4.3}
\end{equation}
The results for these conditions are shown in Fig.\,\ref{f4}. They are
similar to the case of the vacuum conditions but with different slopes
in power law dependencies on the (large) Hall number. The power index in
the dependence of the growth rate changes to $3/4$ and the absolute
value of the index for the wave number reduces about two times.

We proceed to analytical estimations of the instability to justify and
partly explain the numerical results.
\subsubsection{Asymptotic estimation}\label{Ass}
The method of asymptotic analysis of resistive hydromagnetic
instabilities was developed in the seminal paper by \citet{FKR63}. We
apply this method to the tearing mode of the Hall instability at large
$R_\mathrm{H}$.

It is convenient to reformulate Eq.\,(\ref{3.4}) in terms of new
variables $\gamma = \sigma/R_\mathrm{H}$ and $v = \mathrm{i}k b/\gamma$:
\begin{eqnarray}
    &&\gamma \left(a + B v\right) = \frac{1}{R_\mathrm{H}}
    \left(\frac{\partial^2 a}{\partial x^2} - k^2 a\right) ,
    \nonumber \\
    &&\gamma^2 v = -k^2\left[ B\left(\frac{\partial^2 a}{\partial x^2}
    - k^2 a\right) - \frac{\partial^2B}{\partial x^2} a\right] +
    \frac{\gamma}{R_\mathrm{H}}\left(\frac{\partial^2 v}{\partial x^2}
    - k^2 v\right) .
    \label{4.4}
\end{eqnarray}
The case of the large Hall number $R_\mathrm{H} \gg 1$ is considered.

\begin{figure}
    \centerline{\includegraphics[width = 9 truecm]{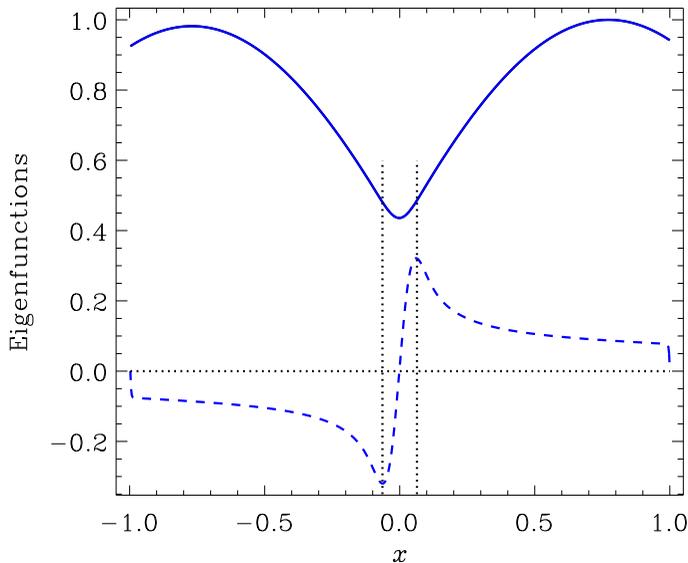}}
    \caption{Profiles of the $x$-component of the current velocity
        $\gamma v(x)$ ({\sl dashed}) and $a(x)$ ({\sl full line}) of the tearing-type
        eigenmode for $R_\mathrm{H} = 10^3$. The vertical dotted lines
        show the central diffusive layer.}
    \label{f5}
\end{figure}

The eigenmodes of Fig.\,\ref{f2} show a thin reconnection layer in the
central part of the slab and smooth variations outside this layer. These
two distinct regions are visualized more clearly by the eigenfunction
profiles of Fig.\,\ref{f5}. The recipe by \citet{FKR63} is to solve for
the (central) diffusive and (outer) non-diffusive regions separately.
Dispersion relation for the growth rates then results from a link of
these solutions.

The unstable eigenmodes grow with \lq hybrid' rates that are large
compared with the rate of diffusion but small compared to the helicoidal
oscillation frequency,
\begin{equation}
    R_\mathrm{H}^{-1} \ll \gamma \ll 1.
    \label{4.5}
\end{equation}
With these inequalities, Eqs.\,(\ref{4.4}) for the non-diffusive regions
reduce to
\begin{eqnarray}
    && v =- a/B
    \nonumber \\
    && \frac{\partial^2 a}{\partial x^2}
    - \left(\frac{1}{B}\frac{\partial^2 B}{\partial x^2} + k^2\right) a = 0.
    \label{4.6}
\end{eqnarray}
The solution of this linear equation for the profile (\ref{4.2}) and
$k\leq \upi/2$ includes two arbitrary constants,
\begin{equation}
    a(x) = C\left(\sin(p x) + C_1\cos(p x)\right) ,
    \label{4.7}
\end{equation}
where $p = \sqrt{\mid\upi^2/4 - k^2\mid }$. The constant $C_1$ can be defined
with boundary conditions that have to be applied separately for positive
and negative $x$. The vacuum boundary condition (\ref{3.5}) for positive
$x$ gives
\begin{equation}
    C_1 = C_+ = \frac{k\sin(p) + p\cos(p)}{p\sin(p) - k\cos(p)},
    \label{4.8}
\end{equation}
while for negative $x$ it is $C_1 = C_- = -C_+$. The constant $C$ in the
\lq external' solution (\ref{4.7}) remains indefinite in the linear
stability problem. However, the constant does not contribute the
stability index
\begin{equation}
    \Delta' = \frac{1}{a}\frac{\partial a}{\partial x}\mid_{x = +0} -
    \frac{1}{a}\frac{\partial a}{\partial x}\mid_{x = -0}
    \label{4.9}
\end{equation}
of \citet{FKR63}. It is accounted for in Eq.\,(\ref{4.9}) that the
logarithmic derivatives of the external solution vary little with $\mid x\mid
\ll 1$ on either side of the origin and their values at the borders of
the diffusive layer can be replaced by the values at the origin.

With the Eq.\,(\ref{4.8}) for the vacuum boundary conditions we get
\begin{equation}
    \Delta' = 2 p \frac{p\sin(p) - k\cos(p)}{k\sin(p) + p\cos(p)} .
    \label{4.10}
\end{equation}
This expression changes from positive to negative values with $k$ increasing beyond the value of about $k = 1.263$. We shall see that the negative values mean stability (note that the wave numbers of Fig.\,\ref{f1} correspond to positive $\Delta'$). Repeating the derivations for $k > \upi /2$ gives the (negative) quantity
\begin{equation}
    \Delta' =-2p\frac{(k + p)\exp(2p) + k - p}
    {(k + p)\exp(2p) - k + p} ,
    \label{4.101}
\end{equation}
to which the $\Delta'$ of Eq.\,(\ref{4.10}) transforms continuously when $k$ exceeds $\upi/2$.

Repeating the derivations for periodic conditions of Eq.\,(\ref{4.3}) we
find positive
\begin{equation}
    \Delta' = 2 p \tan(p)
    \label{4.11}
\end{equation}
for $k \leq \upi/2$ and negative $\Delta' = -2p[\exp(2p) - 1]/[\exp(2p) + 1]$ for $k > \upi/2$.

It can be shown that $\Delta'$ for perfect conductor
boundary conditions or the linear profile of $B$ is negative for any $k$ thus confirming numerically found stability for these cases.

Inside the \lq inner' diffusive region, spatial derivatives are not
small $\partial /\partial x \gg 1$. The region is also thin so that the
the background field can be approximated by the linear profile $B(x) =
xB'$, where the prime means the derivative at $x = 0$. Eqs.\,(\ref{4.4})
then reduce to
\begin{eqnarray}
    &&\gamma \left( a - \frac{\upi}{2} x v\right) =
    \frac{1}{R_\mathrm{H}}\frac{\partial^2 a}{\partial x^2}
    \nonumber \\
    && \gamma^2 v = k^2 x \frac{\upi}{2}\frac{\partial^2 a}{\partial x^2}
    + \frac{\gamma}{R_\mathrm{H}}\frac{\partial^2 v}{\partial x^2} ,
    \label{4.12}
\end{eqnarray}
where $B' = -\upi /2$ for the profile (\ref{4.2}) was used; a
generalization to any other antisymmetric profile is easy.

An expression for the jump in the logarithmic derivative across the
diffusion layer
\begin{equation}
    \Delta(\gamma) = \frac{1}{a}\frac{\partial a}{\partial x}\mid_{x = \delta} -
    \frac{1}{a}\frac{\partial a}{\partial x}\mid_{x = -\delta}
    \label{4.13}
\end{equation}
should be derived from Eqs.\,(\ref{4.12}) to get the dispersion relation
$\Delta(\gamma) = \Delta'$.

The $\Delta(\gamma)$ is derived in the so-called \lq constant $\psi$
approximation' \citep[][Sect.\,6.8.1]{FKR63,P14} in Appendix\,\ref{A}.
Eq.\,(\ref{a17}) leads to the following estimation for the growth rate
\begin{equation}
    \sigma = R_\mathrm{H}\left( - k + \sqrt{k^2 +
    1.18\Delta'(k)\,k^{3/2}R_\mathrm{H}^{-1/2}}\right),
    \label{4.14}
\end{equation}
where positive $\Delta'$ only result in positive growth rates.
The growth rates estimated with this equation are shown by the dashed
lines in Figs.\,\ref{f1} and \ref{f4} for the vacuum and periodic
boundary conditions respectively. The $\Delta'$ of the corresponding
equations (\ref{4.10}) and (\ref{4.11}) were used in the estimations
with the wave numbers taken from the corresponding figures.
In contrast with the numerical computations, the asymptotic estimations
apply to large Hall numbers only but are not restricted to the
moderately large $R_\mathrm{H}$.

The analytic and numerical trends are similar and they approach each
other with increasing Hall number. The asymptotic theory overestimates
the numerical growth rates by about 20\% for the largest Hall numbers in
Fig.\,\ref{f1} and by almost 30\% in Fig.\,\ref{f4}. The difference can
be related to the approximate nature of the asymptotic theory. The clear
similarity of the analytic and numerical results nevertheless confirms
the numerically detected dependence of the Hall instability scaling on
boundary conditions. The asymptotic estimations also confirm the
stability of the linear profile of the background field and the
stability for perfect conductor boundary conditions in our model.

\subsection{Symmetric field profiles}
According to the proposed interpretation of the Hall instability, it can
be expected if the condition of Eq.\,(\ref{4.1}) is satisfied. The
expectation is generally confirmed by computations for the parabolic
profile
\begin{equation}
    B(x) = \beta - x^2 .
    \label{4.15}
\end{equation}
In agreement with the condition (\ref{4.1}), instability was found for
positive $\beta$ only. It can be expected that the local condition
should be satisfied in a sufficiently broad region and the parameter
$\beta$ should be sufficiently large for the onset of the instability.
The expectation is generally confirmed by computations, but the critical
value $\beta_\mathrm{c}$ for the instability is rather small and
decreases with the Hall number: $\beta_c \approx 5\times 10^{-3}$ for
$R_\mathrm{H} = 10^3$ and $\beta_c \approx 3\times 10^{-4}$ for
$R_\mathrm{H} = 10^5$. The condition (\ref{4.1}) is fulfilled in the region of $-\beta^{1/2} < x < \beta^{1/2}$.

\begin{figure}
    \centerline{\includegraphics[width = \columnwidth]{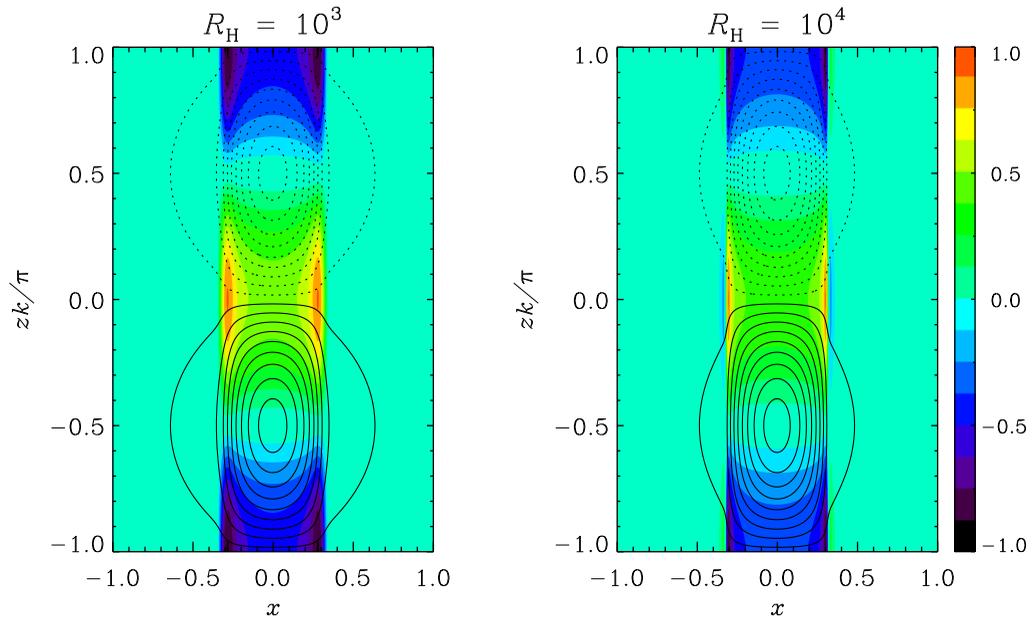}}
    \caption{Poloidal field lines and isolines of the toroidal field of most rapidly
        growing instability modes for $R_\mathrm{H} = 10^3$ ({\sl left panel}) and
        $R_\mathrm{H} = 10^4$ ({\sl right panel}). Full and dashed lines correspond
        to the clockwise and anti-clockwise circulation of the poloidal field
        vector respectively. All for the symmetric field profile of
        Eq.\,(\ref{4.15}) with $\beta = 0.1$.}
    \label{f6}
\end{figure}

Figure\,\ref{f6} shows that unstable eigenmodes are confined to this region.

\begin{figure}
    \centerline{\includegraphics[width = \columnwidth]{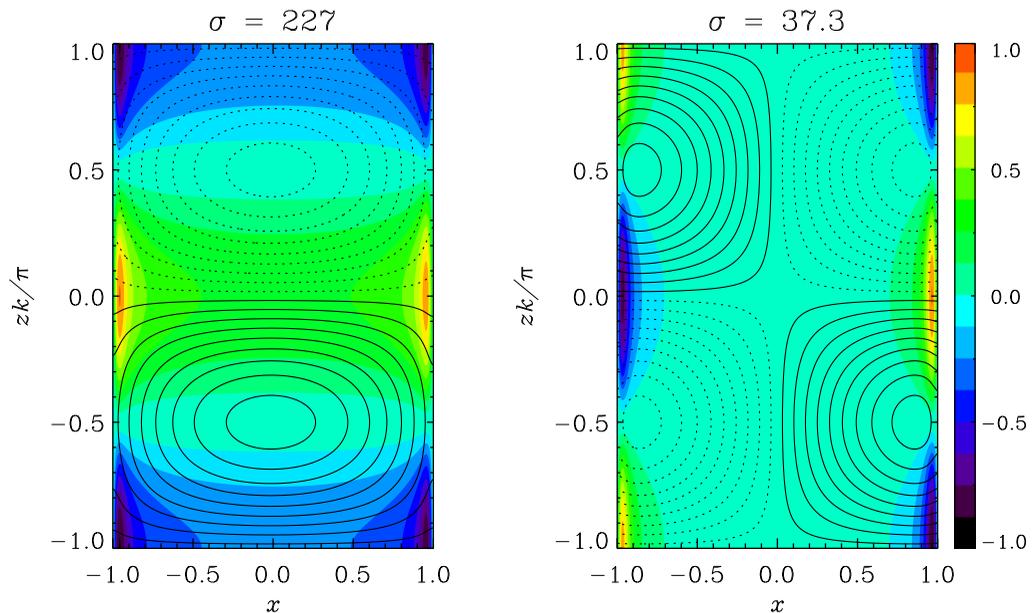}}
    \caption{Poloidal field lines and isolines of the toroidal field for
        the symmetric ({\sl left panel}) and antisymmetric ({\sl right panel})
        instability modes. The corresponding normalized growth rates are
        given at the top. All for the symmetric profile of
        Eq.\,(\ref{4.15}) with $\beta = 1$ and $R_\mathrm{H} = 10^3$.}
    \label{f7}
\end{figure}

The modes of this Figure do not belong to the tearing type but possess a
fine structure near the border of the instability region. The region
broadens with increasing $\beta$ and occupies the entire slab for $\beta
= 1$ (Fig.\,\ref{f7}). Another difference from the antisymmetric field
profile, for which only one unstable mode was found for a given $k$, is
that two unstable modes symmetric and antisymmetric about the slab
midplane can exist for the symmetric profile of Eq.\,(\ref{4.15}). These
two types of symmetry are illustrated by Fig.\,\ref{f7}. The
antisymmetric mode always had a smaller growth rate compared to the
symmetric mode.

\begin{figure}
    \centerline{\includegraphics[width = \columnwidth]{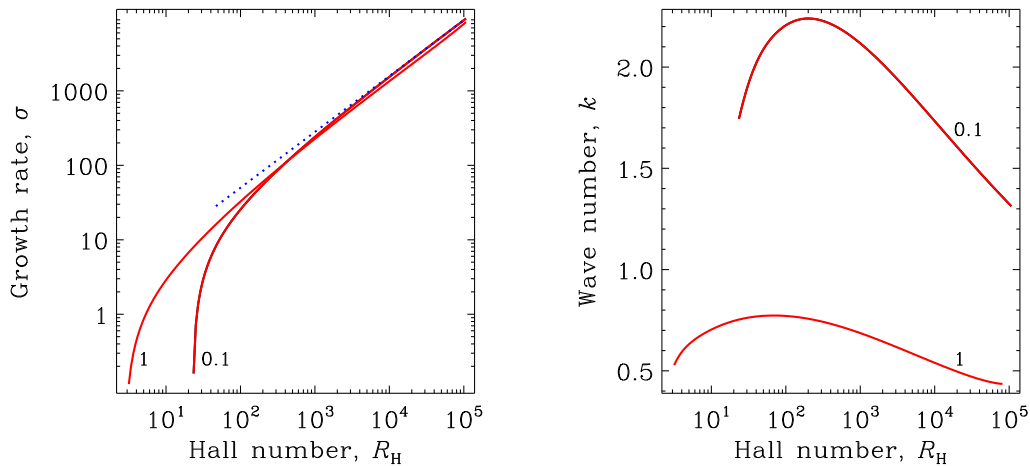}}
    \caption{{\sl Left panel}: Growth rates for the symmetric field profile of
           Eq.\,(\ref{4.15}) with $\beta = 0.1$ and $\beta = 1$.
           The dotted line shows the asymptotic power-law dependence
           $\sigma\propto R_\mathrm{H}^{3/4}$.
           {\sl Right panel}: Wave numbers for which the maximum growth rates
           of the left panel were obtained.
           The lines are marked by the corresponding values of $\beta$.}
    \label{f8}
\end{figure}

The maximum growth rates for $\beta = 0.1$ and $\beta = 1$ are shown in
Fig.\ref{f8} as functions of the Hall number. The growth rates approach
the power law of $\sigma \propto R_\mathrm{H}^{3/4}$ for large Hall
number. The asymptotic growth rates depend weakly on $\beta$ but the
asymptotic trend is approached at larger $R_\mathrm{H}$ for smaller
$\beta$.

The perfect conductor boundary conditions (\ref{3.6}) switch the
instability off for $\beta = 1$ similar to the case of the antisymmetric
field profile. For small $\beta$ however, the instability is weakly
sensitive to the boundary conditions, though the growth rates for
perfect conductor boundaries are somewhat smaller compared to the vacuum
conditions. This is probably because the unstable modes for small
$\beta$ occupy the central part of the slab only (Fig.\,\ref{f6}) and do
not feel the boundaries.

\section{Discussion}
The formerly proposed interpretation of the Hall instability as driven
jointly by the helicoidal rotation and shear in the current velocity
\citep{Kit17} is generally confirmed by the computations of this paper.
According to this interpretation, instability is expected for the
background field profiles having a region of counter-aligned magnetic
field and vorticity of the current velocity. For 1D profiles of this
paper, this means that inequality (\ref{4.1}) should be satisfied for
instability. The inequality is satisfied by the antisymmetric profile
(\ref{4.2}) supporting the tearing-type mode of the instability.
However, the instability is switched off when the antisymmetric linear
profile with uniform current velocity is approached (Fig.\,\ref{f3}).
Figure\,\ref{f6} shows that unstable modes for the symmetric field are
restricted to the region where the inequality (\ref{4.1}) is satisfied.

The proposed interpretation also explains the diffusive character of the
instability. The helicoidal rotation pushes the magnetic field lines in
the direction of decreasing field strength and therefore decreasing rate
of this rotation. The instability in the region of the relatively strong
field is blocked by the less unstable regions of the weaker field. This
\lq displacements jam' is resolved by diffusion. This picture can also
explain the instability switch off by the perfect conductor boundary
conditions: the superconducting walls block helicoidal rotation. The
numerical results for the tearing mode are confirmed by independent
asymptotic estimations. The analytical estimations confirm in particular
the stabilizing effect of superconducting boundaries.

This stabilizing effect is probably not of a general nature. It can be
expected to weaken for \lq sufficiently broad' profiles of the
background field tending to constant far away from the location of the
field reversal. Are the broad profiles realistic however? The Hall
instability can be relevant to the release of magnetic energy from
small-scale magnetic structures. The structures have neither vacuum no
superconducting boundaries but are surrounded by other similar
structures. The periodic conditions of Eq.\,(\ref{4.3}) seem to be more
adequate for this case.

The growth rates approach asymptotically the power law dependence
$\sigma \propto R_\mathrm{H}^\alpha$ on the Hall number with $\alpha =
3/4$; $\alpha = 2/3$ for the case of the tearing mode with vacuum
boundary conditions. This means that the growth rate in physical units
scales as $\sigma \propto B^\alpha\eta^{1-\alpha}$. The growth rate of
the Hall instability increases more steeply with the magnetic field
compared to the hydromagnetic tearing instability with $\alpha = 1/2$.
The Hall instability can produce current sheets without magnetic
reconnection (Fig.\,\ref{f6}). The instability gives an additional
channel for the release of magnetic energy complementary to the
canonical tearing mode. Hall instability catalyzes
resistive dissipation by decreasing spatial scale of magnetic field. It
may be pertinent to note that the Hall number (\ref{2.2}) does not
include any spatial scale. A decrease of the field scale cannot
therefore be balanced by diffusion. Release of magnetic energy on
nonlinear stage of the instability proceeds in a sequence of spikes or
bursts, each spike releases several percent of magnetic energy
\citep{Kit17,Kit19}.

The condition of the counter-aligned magnetic field and vorticity of the
current velocity can be written as
\begin{equation}
    \boldsymbol{B\cdot}\left( \boldsymbol{\nabla}\times(
    \boldsymbol{\nabla}\times\boldsymbol{B})\right) > 0
    \label{5.1}
\end{equation}
for 3D fields. This heuristic condition for the instability is satisfied
by any force-free field of $\boldsymbol{\nabla}\times\boldsymbol{B} =
\alpha(\boldsymbol{r})\boldsymbol{B}$. Instability of a 1D force-free
field with maximum helicity was confirmed in \citet{Kit19}. The
instability can be active with the force-free fields of the solar corona
where the Hall number $R_\mathrm{H} \sim 10^7$ is large.
\bl

This work was financially supported by the Ministry of Science and High
Education of the Russian Federation and by the Russian Foundations for
Basic Research (Project 19-02-00088). The author is thankful to Ms Jennifer
Sutton for language editing.
\bl

Declaration of Interests. The author reports no conflict of interest.

\appendix\section{Derivation of $\Delta (\gamma)$}\label{A}
Equations (\ref{4.12}) for the diffusive layer can be simplified with
the redefined variables $\tilde{v} = \upi v/2,\ \tilde{\gamma} =
2\gamma/(\upi k),$ and $\tilde{R} = k\upi R_\mathrm{H}/2$:
\begin{eqnarray}
    &&\frac{\partial^2 a}{\partial x^2} = \tilde{R}\tilde{\gamma}\left(
    a - x\tilde{v}\right) ,
    \label{a1}
    \\
    &&\tilde{\gamma}^2\tilde{v} = x\frac{\partial^2 a}{\partial x^2} +
    \frac{\tilde{\gamma}}{\tilde{R}}\frac{\partial^2\tilde{v}}
    {\partial x^2}.
    \label{a2}
\end{eqnarray}
Integration of Eq.\,(\ref{a1}) across the diffusive layer yields
the $\Delta$-parameter of Eq.\,(\ref{4.13}),
\begin{equation}
    \Delta =
    \frac{\tilde{\gamma}\tilde{R}}{a(\delta)}\int_{-\delta}^{\delta}
    \left(a(x) - x\tilde{v}(x)\right) \mathrm{d} x ,
    \label{a3}
\end{equation}
where $\delta \ll 1$ is a small distance outside of the diffusive layer and it is taken into account that $a(x)$ profile is symmetric with $a(\delta) = a(-\delta)$ (Fig.\,\ref{f5}).

Substitution of $\partial^2 a/\partial x^2$ from Eq.\,(\ref{a1}) into
Eq.\,(\ref{a2}) gives the differential equation
\begin{equation}
    \frac{\partial^2\tilde{v}}{\partial s^2} - s^2\tilde{v} -
    \tilde{\gamma}\tilde{v} = -\tilde{R}^{1/2} s a ,
    \label{a4}
\end{equation}
in terms of a new variable $s = \tilde{R}^{1/2} x$

The idea with the constant $\psi$ approximation - constant $a$ in the
present case - is to express $\tilde{v}$ in terms of $a$ using the
Eq.\,(\ref{a4}) and then put $a$ outside the integral of Eq.\,(\ref{a3})
based on the fact that it varies little inside the diffusive layer
(Fig.\,\ref{f5}). This excludes the (still unknown) constant $a$ from
the expression for $\Delta$ and gives the expression in terms of the
parameters $\gamma$, $R_\mathrm{H}$ and $k$, as desired.

Applying this approximation to Eq.\,(\ref{a3}) and using the new variable $s$ gives
\begin{equation}
    \Delta =
    \tilde{R}^{1/2}\tilde{\gamma}\int_{-\tilde{R}^{1/2}\delta}^{\tilde{R}^{1/2}\delta}
    \left(1 - \frac{s}{a\tilde{R}^{1/2}}\tilde{v}(s)\right) \mathrm{d} s .
    \label{a5}
\end{equation}
It may be noted that each of the two terms of the integral in this expression diverge with increasing Hall number but the two diverging terms almost balance each other to combine into a finite quantity.

The operator on the left-hand side of Eq.\,(\ref{a4}) has the Hermite
functions,
\begin{equation}
    \varphi_n(s) = \frac{(-1)^n}{\upi^{1/4}\sqrt{2^n n\mathrm{!}}}\mathrm{e}^{s^2/2}
    \frac{\mathrm{d}^n}{\mathrm{d}s^n}\mathrm{e}^{-s^2} ,
    \label{a6}
\end{equation}
as its eigenfunctions:
\begin{equation}
    \frac{\mathrm{d}^2\varphi_n}{\mathrm{d} s^2} - s^2\varphi_n =
    -(2n + 1)\varphi_n.
    \label{a7}
\end{equation}
The functions make up a complete (orthogonal and normalized) basis that
can be used for solving the equation (\ref{a4}) in terms of the series
expansion
\begin{equation}
    \tilde{v}(s) = a\,\tilde{R}^{1/2}\sum_{n = 0}^\infty
    \frac{\varphi_n(s)}{2n + 1 + \tilde{\gamma}}
    \int_{-\infty}^\infty \varphi_n (s') s'
    \mathrm{d} s' .
    \label{a8}
\end{equation}
Note that the solution for not constant $a$ can be obtained by placing
$a$ under the integral sign on the right-hand side of Eq.\,(\ref{a8}).

Expansion of the entire integrand in Eq.\,(\ref{a5}) in terms of the
Hermite functions and using Eq.\,(\ref{a8}) give
\begin{equation}
    \Delta = \tilde{R}^{1/2}\tilde{\gamma}\sum_{n = 0}^\infty
    \left[ \left(\int_{-\infty}^\infty\varphi_n(s)\mathrm{d} s\right)^2
    - \frac{1}{2n + 1 + \tilde{\gamma}}
    \left(\int_{-\infty}^\infty s \varphi_n(s)\mathrm{d}
    s\right)^2\right] ,
    \label{a9}
\end{equation}
where the integration limits are changed to infinity for the asymptotic
case of $\tilde{R} \gg \delta^{-2}$.

The symmetry relation $\varphi_n(s) = (-1)^n\varphi_n(-s)$ shows that the
first term under the summation sign in Eq.\,(\ref{a9}) is zero for odd
$n$ and the second term is zero for even $n$. The equation can then be
written as
\begin{equation}
    \Delta = \tilde{R}^{1/2}\tilde{\gamma}\sum_{m = 0}^\infty
    \left[ \left(\int_{-\infty}^\infty\varphi_{2m}(s)\mathrm{d} s\right)^2
    - \frac{1}{4m + 3 + \tilde{\gamma}}
    \left(\int_{-\infty}^\infty s \varphi_{2m+1}(s)\mathrm{d}
    s\right)^2\right] ,
    \label{a10}
\end{equation}
and the relations
\begin{equation}
    \int_{-\infty}^\infty s\,\varphi_{2m+1}(s)\,\mathrm{d} s =
    \sqrt{4m + 2}
    \int_{-\infty}^\infty\varphi_{2m}(s)\,\mathrm{d} s
    \label{a11}
\end{equation}
and
\begin{equation}
   \left( \int_{-\infty}^\infty\varphi_{2m}(s)\,\mathrm{d} s\right)^2 =
   2\frac{\Gamma(m + 1/2)}{\Gamma(m + 1)}
   \label{a12}
\end{equation}
can be used to simplify it further:
\begin{equation}
    \Delta = 2 \tilde{R}^{1/2}\tilde{\gamma}\sum_{m = 0}^\infty
    \frac{1 +\tilde{\gamma}}{4m + 3 + \tilde{\gamma}}
    \frac{\Gamma(m + 1/2)}{\Gamma(m + 1)},
    \label{a13}
\end{equation}
where $\Gamma$ is the Gamma function.

As has been noted, each of the two series on the right-hand side of
Eq.\,(\ref{a10}) diverges when summed-up separately, but their
combination of Eq.\,(\ref{a13}) converges ($\Gamma(m + 1/2)/\Gamma(m +1)
\propto m^{-1/2}$ for large $m$). The convergence is however slow and
Eq.\,(\ref{a13}) is difficult to handle. The series in this equation can
be rearranged in order to extract its slowly converging part that can be
summed-up analytically and the remaining rapidly converging part is easy
for numerical estimation:
\begin{equation}
    \Delta = 2\sqrt{\upi\tilde{R}}\,\tilde{\gamma}(1 + \tilde{\gamma})
    \sum_{n = 0}^\infty (-1)^n\tilde{\gamma}^n A_n ,
    \label{a14}
\end{equation}
where
\begin{equation}
    A_n = \sum_{m=0}^\infty\frac{C_m}{(4m + 3)^{(n + 1)}},\ \ C_0 = 1,\
    \mathrm{and}\
    C_m = \frac{2m -1}{2m}\,C_{m-1}\ \mathrm{for}\ m > 0.
    \label{a15}
\end{equation}
The geometric series expansion of the term $1/(4m + 3 + \tilde{\gamma})$ in Eq.\,(\ref{a13}) in powers of $\tilde{\gamma}/(4m + 3)$ served to derive the Eqs\,(\ref{a14}) and (\ref{a15}).
The first four coefficients in the series of Eq.\,(\ref{a14}) read
\begin{equation}
    A_0 = \sqrt{\upi}\,\frac{\Gamma(3/4)}{\Gamma(1/4)} = 0.59907,\
    A_1 = 0.12856,\ A_2 = 0.03897,\ A_3 = 0.01259 ,
    \label{a16}
\end{equation}
where the last three coefficients were evaluated numerically from their
rapidly converging series of Eq.\,(\ref{a15}).

Equation (\ref{a14}) gives the expansion for $\Delta$ in powers of the
asymptotically small parameter $\tilde\gamma$. Cutting the expansion
after the second-order terms gives
\begin{equation}
    \Delta = 0.847\sqrt{\frac{R_\mathrm{H}}{k^3}}\left( \gamma^2 +
    2k\gamma \right) ,
    \label{a17}
\end{equation}
where we returned to the \lq unwaved' parameters of Sect.\,\ref{Ass}.
\bibliographystyle{jpp}
\bibliography{paper}
\end{document}